# Achieving Large, Tunable Strain in Monolayer Transition-Metal Dichalcogenides

Abdollah (Ali) M. Dadgar [1,2], Declan Scullion [3], Kyungnam Kang [4], Daniel Esposito [5], Eui-Hyoek Yang [4], Irving P. Herman [6], Marcos A. Pimenta [7], Elton-J. G. Santos [8], Abhay N. Pasupathy [2]

[1] Department of Mechanical Engineering, Columbia University, New York, NY 10027, USA

[2] Physics Department, Columbia University, New York, NY, 10027, USA

[3] School of Mathematics and Physics, Queens University, Belfast, BT7 1NN, UK

[4] Department of Mechanical Engineering, Stevens Institute of Technology, Castle Point on the Hudson, Hoboken, NJ, USA

[5] Department of Chemical Engineering, Columbia University - New York, NY 10027, USA

[6] Department of Applied Physics and Applied Mathematics, Columbia University, New York - NY 10027, USA

[7] Department of Physics, Universidade Federal de Minas Gerais (UFMG), Brazil

[8] School of Chemistry and Chemical Engineering, Queen's University - Belfast, BT9 5AL, UK

KEYWORDS: Strain, $WSe_2$, $WS_2$, Raman Spectroscopy, Transition Metal Dichalcogenides, TMD, phonon structure, polarized Raman.

**ABSTRACT:** We describe a facile technique based on polymer encapsulation to apply several percent controllable strains to monolayer and few-layer Transition Metal Dichalcogenides (TMDs). We use this technique to study the lattice response to strain via polarized Raman spectroscopy in monolayer $WSe_2$ and $WS_2$. The application of strain causes mode-dependent redshifts, with larger shift rates observed for in-plane modes. We observe a splitting of the degeneracy of the in-plane $E'$ modes in both materials and measure the Grüneisen parameters. At large strain, we observe that the reduction of crystal symmetry can lead to a change in the polarization response of the $A'$ mode in $WS_2$. While both $WSe_2$ and $WS_2$ exhibit similar qualitative changes in the phonon structure with strain, we observe much larger changes in mode positions and intensities with strain in $WS_2$. These differences can be explained simply by the degree of iconicity of the metal-chalcogen bond.

One of the iconic characteristics of monolayer 2D materials is their incredible stretchability which allows them to be subjected to several percent strain before yielding [1]. The application of moderate (~1%) strains is expected to change the anharmonicity of interatomic potentials [2, 3], phonon frequencies [4, 5] and effective masses [6, 7]. At larger strains, topological electronic[8] [9] and semiconductor-metal structural phase changes have been predicted [10-13]. Important technological applications such as piezoelectricity can be explored by the application of systematic strain [14, 15]. One of the chief problems in achieving reproducible strain is the intrinsic nature of 2D materials as single layer sheets - they need to be held to a flexible substrate which is then stretched or compressed. Previous experiments [16-19] have used flexible polymers as substrates and metal or polymer caps in order to constrain the 2D material. Using these techniques, approximate strains up to 4% have been reported so far in the literature, but independent verification of the applied strain has been lacking. Achieving large reproducible strains in engineered geometries will allow us to probe these exciting properties of individual 2D materials and their heterostructures [4, 17, 20-26].

In this work, we develop a new strain platform to apply large range accurate uniaxial tensile strains on monolayer and few-layer materials. One of our chief innovations is the development of a novel polymer-based encapsulation method to enable the application of large strain to 2D materials. Here, we apply this technique to study the strain-dependent properties of monolayer WSe2 and WS2 grown by Chemical Vapor Deposition (CVD) on SiO2/Si substrates [27-29]. We use cellulose acetate butyrate (CAB) to lift the monolayers from the SiO2/Si substrates and transfer to polycarbonate substrates. The two polymers are then bonded to produce encapsulated monolayers and multilayers. The key to achieving good bonding is perfect control over the temperature, time and pressure during the bonding process. Additionally, polymer layers that are in the amorphous phase cause nonlinear strain-deflection behavior which is not desirable in our experiments. To resolve this issue, we crystallize the polymer stacks by annealing near the glass transition temperature followed by slow cooling. The crystallized polymers are fully flexible, elastic and springy substrates as shown in Fig. 1(a). After all of our processing steps, we find that the polymer stacks enter into the plastic regime at 7% strain. We find that strains up to this value are perfectly transferred to the encapsulated 2D material as described below.

Our strain method adopts the extra-neutral axis bending technique – Fig. 1(b) in which areas above the neutral axis undergo tensile strain while those below the axis experience compressive strain. In our method, we use a screw-driven vertical translation stage to apply strain to the polymer stacks. We solve the Euler-Bernoulli equation for our geometry in order to achieve an accurate relation between the vertical displacement $\delta$ of the translation stage and the strain $\varepsilon$ of the 2D material. For a fully isotropic, linear and elastic material, the strain-displacement relation is derived as: $\varepsilon = 3t\delta/a(3b + 2a)$ where $t$ is the substrate thickness, $b$ and $a$ are center support and cantilever lengths respectively. In our experiments, the use of a fine adjustment screw gives us a resolution of 0.05% strain for 0.5 mm substrates, with essentially no limit to the maximum strain that can be applied. More details are provided in the Supplemental Material.

Shown in Fig. 1(c) is an optical image of triangular flakes of $WSe_2$ encapsulated by this process. We adjust the CVD process to produce triangular flakes in order to easily identify the crystallographic directions of the grown monolayers. Since the strains achievable in our experiments are large, we can directly verify from optical measurements that the strain being applied to 2D layer is the calculated value. This is illustrated in Fig. 1(d). Each of these images is obtained by overlaying two images, one at zero strain and one at a fixed value of strain (4.2% and 6.5% respectively). Only the edges of the triangles are shown in the images, which are lined up to be at the same vertical height at the top vertex of the triangle. We can directly see by inspection that the length of the triangle along the strain direction is larger when strained as one expects. A pixel-height measurement of the edge-detected images gives us a direct experimental measure of the applied strain, which can be compared to the calculated strain based on the screw displacement. It is found that the two measurements match within 0.1% absolute strain. Thus, our technique allows for the application of uniform, highly repeatable and independently measurable strain on TMD monolayers and heterostructures.

In order to probe the effects of strain on our samples, we choose to characterize with Raman spectroscopy - a simple yet powerful way to measure lattice properties and their coupling to the electronic degrees of freedom. Strains were applied in both zigzag and armchair directions (Y and X axes in Fig. 1(e) ) in our experiments. Our Raman setup with 532 nm excitation wavelength is shown in Fig. 1(f). The measurements were performed while controlling for the incident light's polarization ($E_i$) direction (θ in Fig. 1(e) ). For each experiment, Raman spectra were collected in both the parallel- ($E_s \parallel E_i$) and cross-polarized ($E_s \perp E_i$) detector geometries, shown with standard notations $Z(YY)\bar{Z}$ and $Z(YX)\bar{Z}$ respectively. In our experiments, we found no dependence of the Raman spectra on the angle of incidence relative to the crystallographic axis at zero strain. We therefore fix our incidence angle to the $Y$ direction, and measure the unpolarized, parallel-polarized and cross-polarized Raman spectra at each value of strain which is applied in the $X$ direction.

We first discuss the properties of monolayer $WSe_2$. Shown in Fig. 2(a) are a sequence of spectra taken at different values of strain in the unpolarized, parallel and cross polarization geometries. Previous Raman spectroscopy measurements performed on monolayer $WSe_2$ have identified three vibrational modes [30-32] termed $A'$, $E'$ and $2LA$. $A'$ is an out-of-plane phonon mode in which the top and bottom chalcogen atoms vibrate in opposing directions; while $E'$ is in-plane mode where the metal atoms vibrate out-of-phase with the chalcogen atoms [33]. The $2LA$ mode results from a double resonance process involving two phonons from the $LA$ branch. Second order processes can in general give rise to a complex lineshape in the Raman spectrum; yet, in the case of $WSe_2$ we find that a single Lorentzian can be used to model well the $2LA$ mode lineshape. Although $A'$ and $E'$ modes are nearly degenerate, they can be distinguished from each other by polarization dependency of their intensities. The out of plane, symmetric $A'$ mode disappears due to its symmetry in the cross polarization geometry, leaving behind only the $E'$ mode. Our spectra in the cross-polarization geometry can thus be modeled well as the sum of two Lorentzian peaks corresponding to $E'$ and $2LA$ modes. Information of the $E'$ mode position can then be used to fit the spectra seen in the parallel polarization geometry in order to extract the nearly-overlapping $A'$ mode position.

Having understood the polarization-dependent Raman spectra of unstrained monolayer $WSe_2$, we apply uniaxial strains and measure the Raman response. The effects of uniaxial strain up to 1% on monolayer $WSe_2$ has previously been experimentally investigated via unpolarized Raman [17] and absorption spectroscopy [34]. Raman spectra under increasing uniaxial strain up to 3% are shown in Fig. 2(a). A close examination of spectral lineshapes in the cross polarization geometry shows that the $E'$ mode becomes broader with increasing strain. In general, we expect that the initially doubly degenerate $E'$ mode splits on the application of strain into $E'^+$ and $E'^-$. The displacement eigenvector of the $E'^+$ mode is orthogonal to the direction of strain, while it is parallel for the $E'^-$ mode, as has previously been observed for $MoS_2$ and graphene [3, 16, 21]. While we cannot observe a complete separation of the $E'^+$ and $E'^-$ modes in our data, it is nevertheless straightforward to fit the lineshape to two Lorentzian functions and extract the splitting as a function of strain, as shown in Fig. 2(e). The splitting of the $E'$ mode under tensile strain due to the anharmonictiy of molecular potentials can be described by Grüneisen parameter $\gamma = (|\Delta\omega_{E'^+}| + |\Delta\omega_{E'^-}|)/2\omega_{E'}(1 - v)$ and the shear deformation potential $\beta = (||\Delta\omega_{E'^+}| - |\Delta\omega_{E'^-}||)/2\omega_{E'}(1 + v)$ where $\omega_{E'}$ is the frequency of $E'$ mode, $\Delta\omega_{E'^+}$ and $\Delta\omega_{E'^-}$ are the frequency shifts of split modes per unit percent strain and $v$ is Poisson's ratio which is 0.27 for our substrates. We obtain values of $\gamma = 0.38$, $\beta = 0.10$ for $WSe_2$ which are smaller than those reported for graphene [2, 3]. Using the fits for $E'$ mode from the cross-polarization geometry, we then extract the behavior of the $A'$ mode as a function of strain. We find that this mode expectedly does not shift significantly with strain due to its nature as an out-of-plane excitation. Finally, the $2LA$ mode can be easily fit in both parallel and cross polarization spectra. Small redshifts are observed in its position with increasing strain. We also observe an increase in the width of the $A'$ and $2LA$ modes with increasing strain, as well as a decrease in the intensities of all observed modes at higher strains. These observations are summarized in Fig. 2(d).

The spectra of $WS_2$ show additional structure [35], as can be seen in Fig. 3(a). Firstly, two additional lines that we term $P_1$ and $P_2$ are observed in the spectra at 303 and 332 cm$^{-1}$ respectively. We clarify the mode assignment of these peaks based on theory as described below. The $A'$ mode is well separated from the other modes and is located at 423 cm$^{-1}$. Secondly, the region between 345 cm$^{-1}$ and 365 cm$^{-1}$ shows a complex lineshape. Previous measurements have shown that the 2$LA$ and $E'$ modes are nearly degenerate in WS$_2$. However, we find that even two Lorentzians are not sufficient to accurately model the lineshape at zero strain, and we need a minimum of three Lorentzians to reasonably fit the lineshape. One of these three peaks is related to $E'$ mode while we associate the other two with the $LA$ branch and label them by 2$LA_1$ and 2$LA_2$. The description of the $LA$ branch as the sum of two Lorentzians has been made before in the case of MoS2 [36]. Such complex lineshapes are in general expected in double-resonance processes where one has to properly account for the phonon density of states as well as the electron-phonon couplings at different points in the Brillouin Zone. Our strain-dependent measurements help make the distinction between the $E'$ and 2$LA$ bands as described below.

Shown in Fig. 3(a) are a sequence of Raman spectra of WS$_2$ obtained at different values of strains. At each strain, we fit both the parallel and cross-polarization spectra to obtain a consistent set of peaks. Similar to WSe$_2$, we find a splitting of the $E'$ mode with applied strain. The measured Grüneisen parameter and shear deformation potential are $\gamma = 0.54$, $\beta = 0.14$ respectively. All extracted peaks positions, intensities and widths as a function of strain are shown in Fig. 3(b,c,d). Similar to WSe$_2$, the $A'$ mode shows a minimal response to strain due to its out of plane nature. Different from WSe$_2$, the intensity of all the modes increases as a function of strain. Very interestingly, we find that at high strain (>2.5%), the $A'$ mode appears in cross polarization geometry while the P1 mode continues to be fully suppressed. We discuss these observations below.

To better understand the vibrational properties of monolayer WS$_2$ and WSe$_2$, we perform first-principles functional theory (DFT) calculations. In order to apply uniaxial strain, we use orthorhombic unit cell of different sizes, where a specific direction of the lattice, either armchair or zigzag is strained. Similar results were obtained at both directions, using a hexagonal cell (see figure S3 in Supplemental Material). Fig. 4(a,b) show the phonon dispersion curves for WS$_2$ and WSe$_2$ respectively at zero and finite strains. Vibrational modes are labelled accordingly to their symmetry at the Γ-point. It can be seen that there are six optical modes (2$E''$, 2$E'$, $A_1'$, $A_2'$) which are consistent with the 2H phase D$_{3h}$ space group symmetry. The observed phonon dispersion is consistent with previous theoretical studies [5, 10, 37-43]. With the application of uniaxial strain we observe a splitting of the doubly degenerate $E'$ and $E''$ modes in both materials, due to reduction of the crystal symmetry. The variation in the frequency of the Raman active modes with applied strain is plotted in Fig. 4(c,d). All modes red shift with uniaxial strain with the exception of the $A'$ mode in WSe$_2$ which shows a variation of 0.03 cm$^{-1}$ per percent strain (see figures S4 and S5 in Supplemental Material). The variations in mode frequency for WS$_2$ are consistent with the experimental results shown in Fig. 3(b). We also calculated the Grüneisen parameters from the phonon dispersion relations at different k-points using $\gamma = -(V/\omega(qv))(\partial\omega(qv)/\partial V)$ (see figure S6 in the Supplemental Material). The results obtained for the $E'$ modes are 0.45 and 0.54 for WSe$_2$ and WS$_2$, respectively, which are in remarkable agreement with the experimental results. Interestingly, these $\gamma$ values are well reproduced using a simple GGA approximation within the DFT functional. This is to be contrasted with other 2D materials such as hBN, where the exchange-correlation potential has to be fine-tuned to achieve a quantitative agreement with experiment [44].

To study the second order Raman active modes, we examine phonon dispersions at the edges of Brillouin Zone. The frequencies of peaks $P_1$ and $P_2$ are found to be consistent with the transverse optical (TO) $E''$ and $E'$ modes respectively [5]. To approximate the frequency of the 2$LA$ mode we multiply the frequency of the $LA$ mode at the M-point of the Brillouin Zone by a factor of two, as this is a two-phonon process. An overall good agreement is seen between experimental and theoretical results with the exception of the 2$LA(M)$ mode of WSe$_2$, which is increasing in wavenumber whereas experimental results show it to be decreasing. The smaller variation of this mode in the calculations is probably due to limitation in calculation of the second-order process.

In general, the intensity of a phonon mode in Raman spectrum depends both on the optical response of the material via the polarizability as well as the details of the electronic structure and electron-phonon coupling. All of these quantities evolve with applied strain. In our theoretical calculations, we examine the contribution of polarizability change with strain to the Raman intensity. In Fig. 4(e,f), the intensities of $A'$ and $E'$ modes show large increase in WS$_2$, and small reduction in WSe2 instead. Both of these are consistent with the experimental observations. Our results for mode frequencies, anharmonicities and polarizabilities give us insight into the nature of the chemical bond between the metal and chalcogen atoms. WS2 has larger phonon frequency by about factor of two with higher anharmonicities, that would be simply expected for lighter chalcogen atom. However, we also see from experiment that its polarizability is several times larger than that of WSe2. This can be explained by comparing the ionic character of the W-S bond with that of the W-Se bond. A Bader analysis performed for both systems with various strains shows that the charge transfer towards the chalcogenide atoms in the W-S and W-Se bonds is -0.80 electrons/unit cell, and -0.18 electrons/unit cell, respectively, at 0% strain. These magnitudes increase by 3.5% and 16.07% as 3% strain applied into WS2 and WSe2, respectively. This difference in the ionic character of the two bonds and their response to strain provides a simple picture for the evolution of Raman spectra in these compounds with strain.

An intriguing finding in our Raman spectra is the presence of the $A'$ mode in the cross-polarization geometry at high strain shown in Fig. 3(a). From the symmetry perspective, the $A'$ mode is not cross-polarized Raman active at zero strain when the unit cell is hexagonal. However, upon the application of strain the symmetry is lowered from hexagonal to moncclininc (for a generic strain direction), making the $A'$ mode observable in the Raman spectrum (see Supplemental Material). In general, we expect the intensity of this mode to increase with the size of the monoclinic distortion which is proportional to applied strain. We investigate this in our theoretical calculations applied to the cross polarization geometry with the monoclinic unit cell. At zero strain we indeed find a complete suppression of the $A'$ mode. At a strain of 3%, we observe a non-zero intensity for the $A'$ mode, though its intensity is 99% lower in cross polarization when compared to the parallel polarization case. These results point to the role of large strains in actually modifying the symmetry of the lattice and thus changing selection rules. Such "strain engineering" is a promising avenue to tune the optoelectronic properties of both the semiconducting and metallic monolayer transition metal dichalcogenides.

**ACKNOWLEDGMENT**

This work is supported by the Air Force Office of Scientific Research (grant number FA9550-16-1-0601, A.M.D.) and by the National Science Foundation (grant number DMR-1610110, A.N.P.). Shared characterization facilities were provided by the Materials Research Science and Engineering Center (provided through the NSF Grant: DMR-1420634). E.H.Y acknowledges NSF grant: ECCS-1104870, and Air Force Office for Scientific Research, grant: FA9550-12-1-0326. M.A.P thanks the Brazilian agencies Fapeming and CNPq for financial support. D.S thanks the studentship from the EPSRC-DTP award. E.J.G.S acknowledges the use of computational resources from the UK national high performance computing service, ARCHER, for which access was obtained via the UKCP consortium and funded by EPSRC grant ref EP/K013564/1; and the Extreme Science and Engineering Discovery Environment (XSEDE), supported by NSF grants number TG-DMR120049 and TG-DMR150017.

We thank Drew Edelberg, Nathan Finney and Nathan Zhao for experimental help, and James Hone and Jeffrey Kysar for discussions.

**Corresponding Author**

* Abhay N. Pasupathy: apn2108@columbia.edu

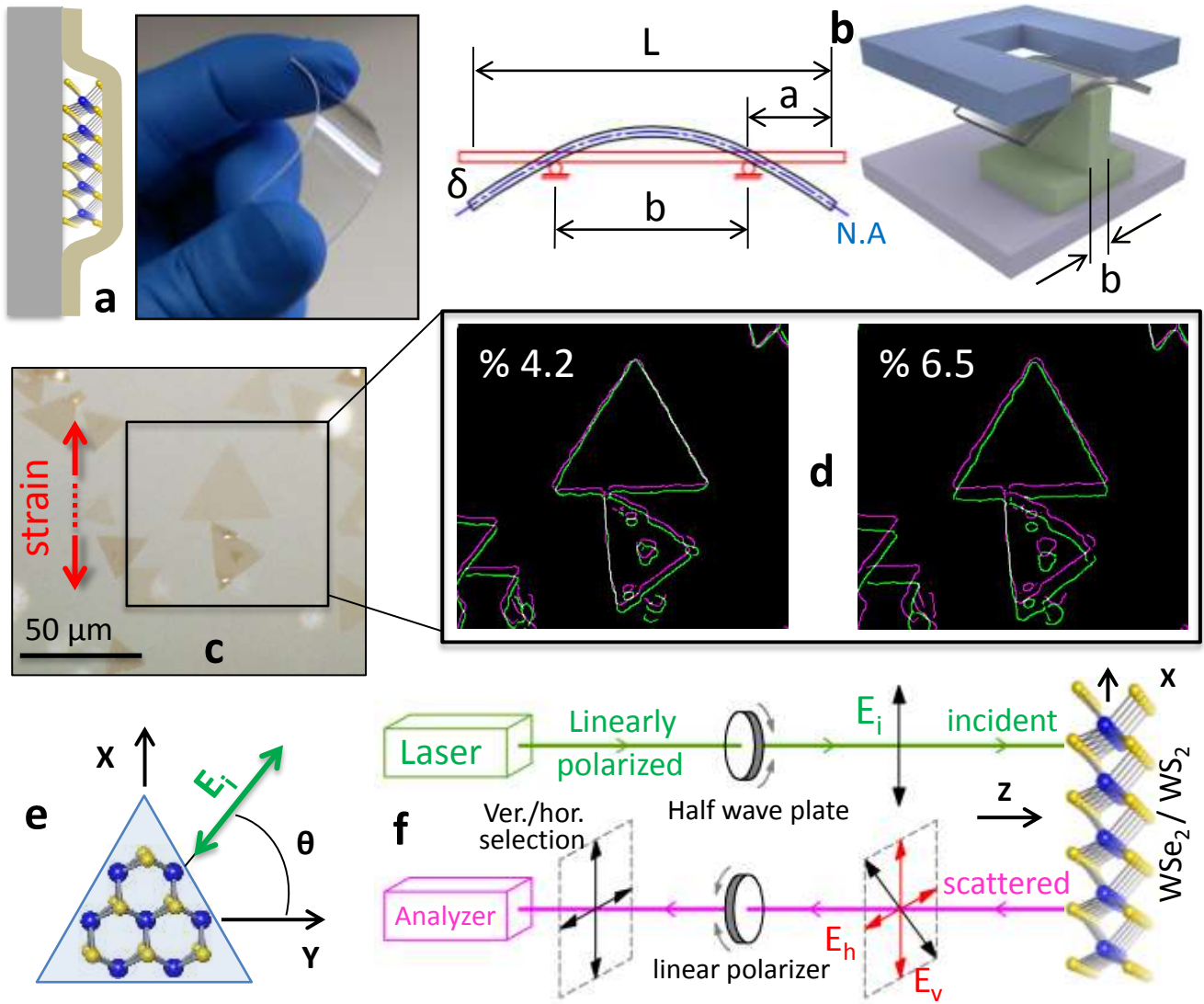

**Figure 1** (a)Polymer encapsulated monolayer TMDs. (b)Strain apparatus. (c)Encapsulated WSe2 monolayers. (d)Overlaid edge-detected images of strained(green) and unstrained(purple) monolayer edges for 4.2% and 6.5% calculated strains. (e)Incident light(Y, $\theta = 0$) and strain(X) directions with respect to crystal lattice. (f)Raman spectrometer setup.

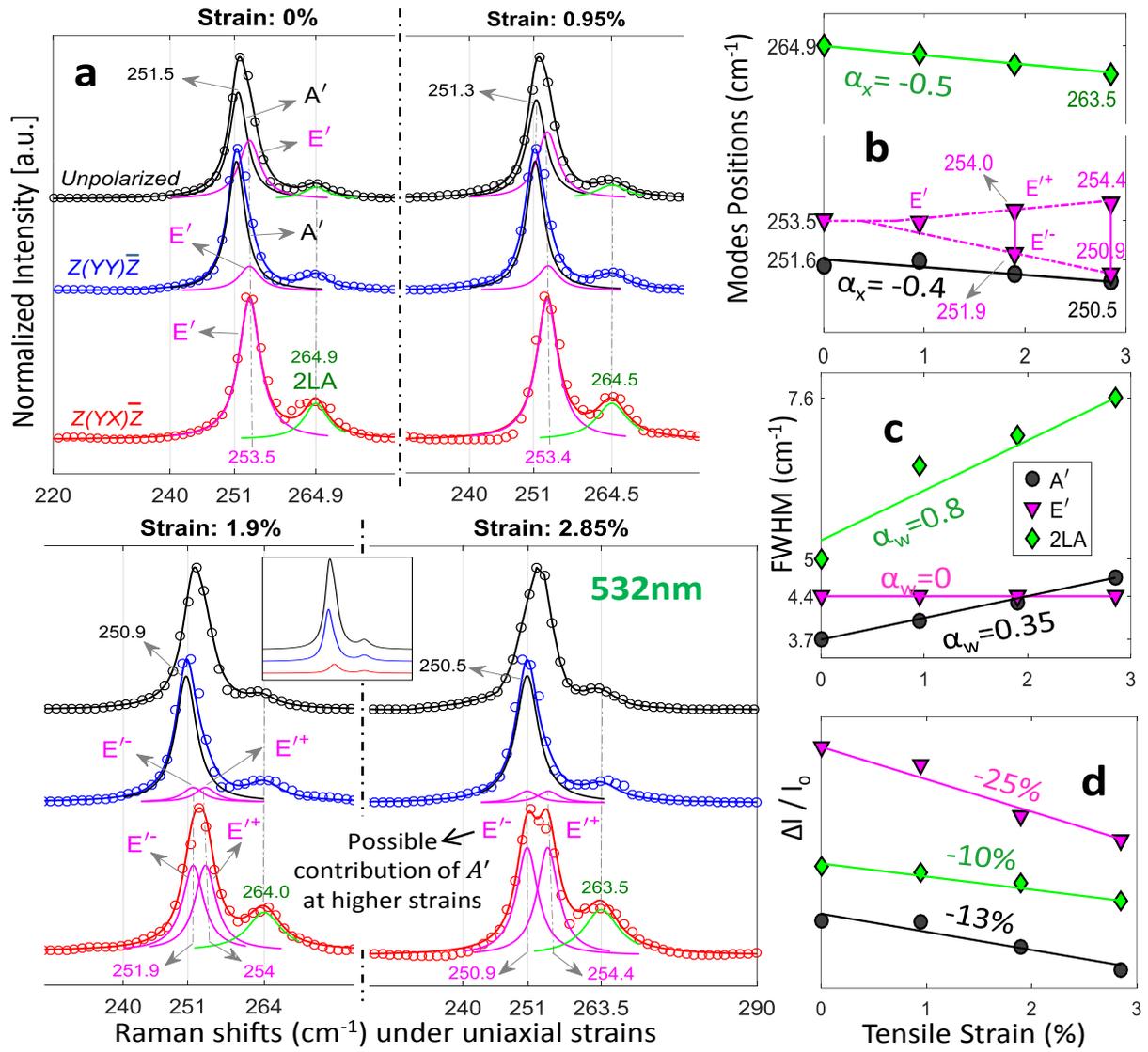

**Figure 2** (a)Unpolarized(black), parallel-polarized(blue) and cross-polarized(red) Raman spectra of monolayer $WSe_2$ under various strains. The spectra are normalized to their peak intensities and shifted along y-axis for better illustration. Inset: real intensities of measured spectra without normalization. (b,c,d)Positions of phonon modes, Full Width at Half Maximum and relative intensity changes vs tensile strain. ±α denotes the amount of change per percent tensile strain.

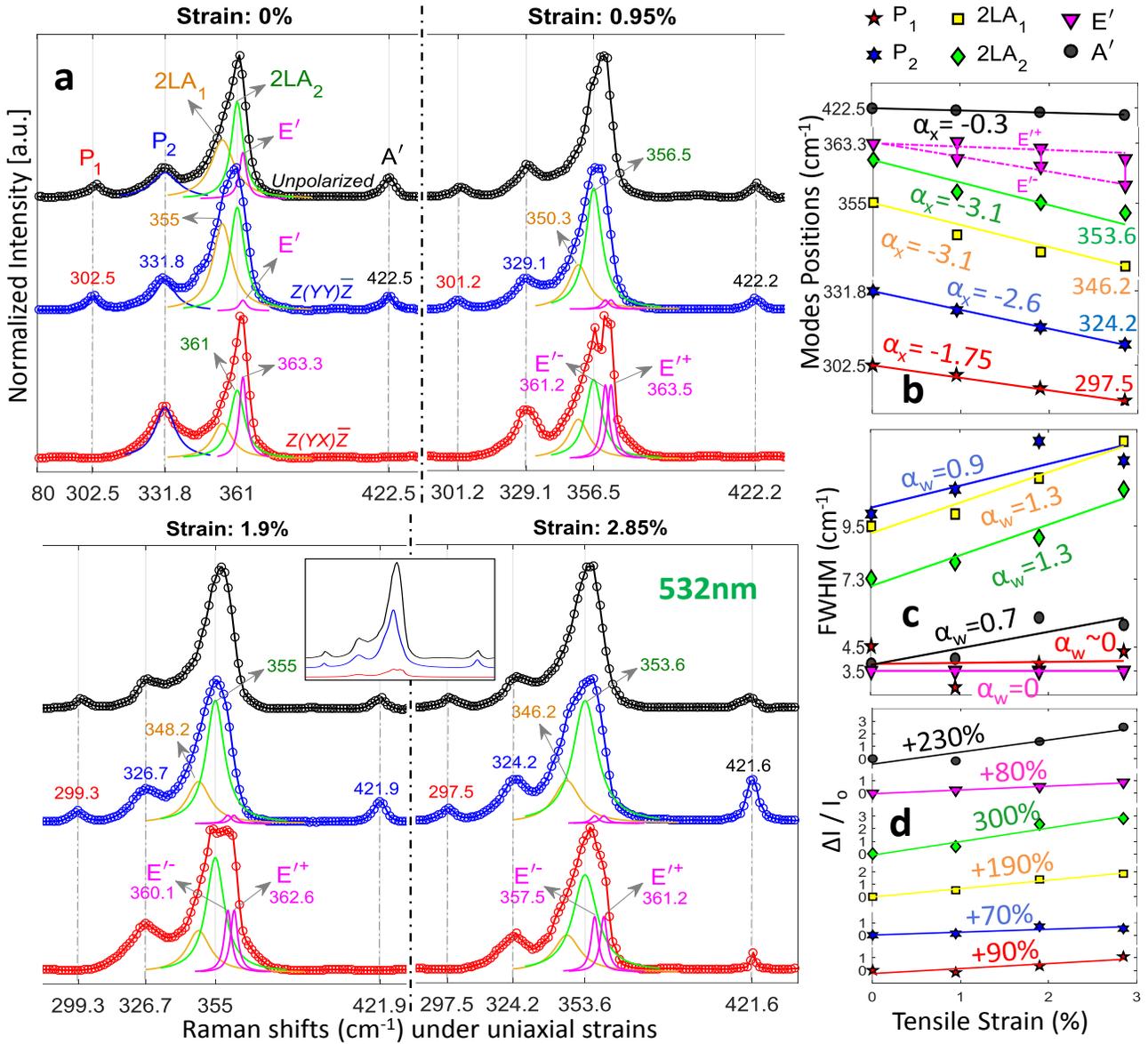

**Figure 3** (a)Unpolarized(black), parallel-polarized(blue) and cross-polarized(red) Raman spectra of monolayer WS$_2$ under various strains. The spectra are normalized to their peak intensities and shifted along y-axis for better illustration. Inset: real intensities of measured spectra without normalization. (b,c,d)Positions of phonon modes, Full Width at Half Maximum and relative intensity changes vs tensile strain. ±α denotes the amount of change per percent tensile strain.

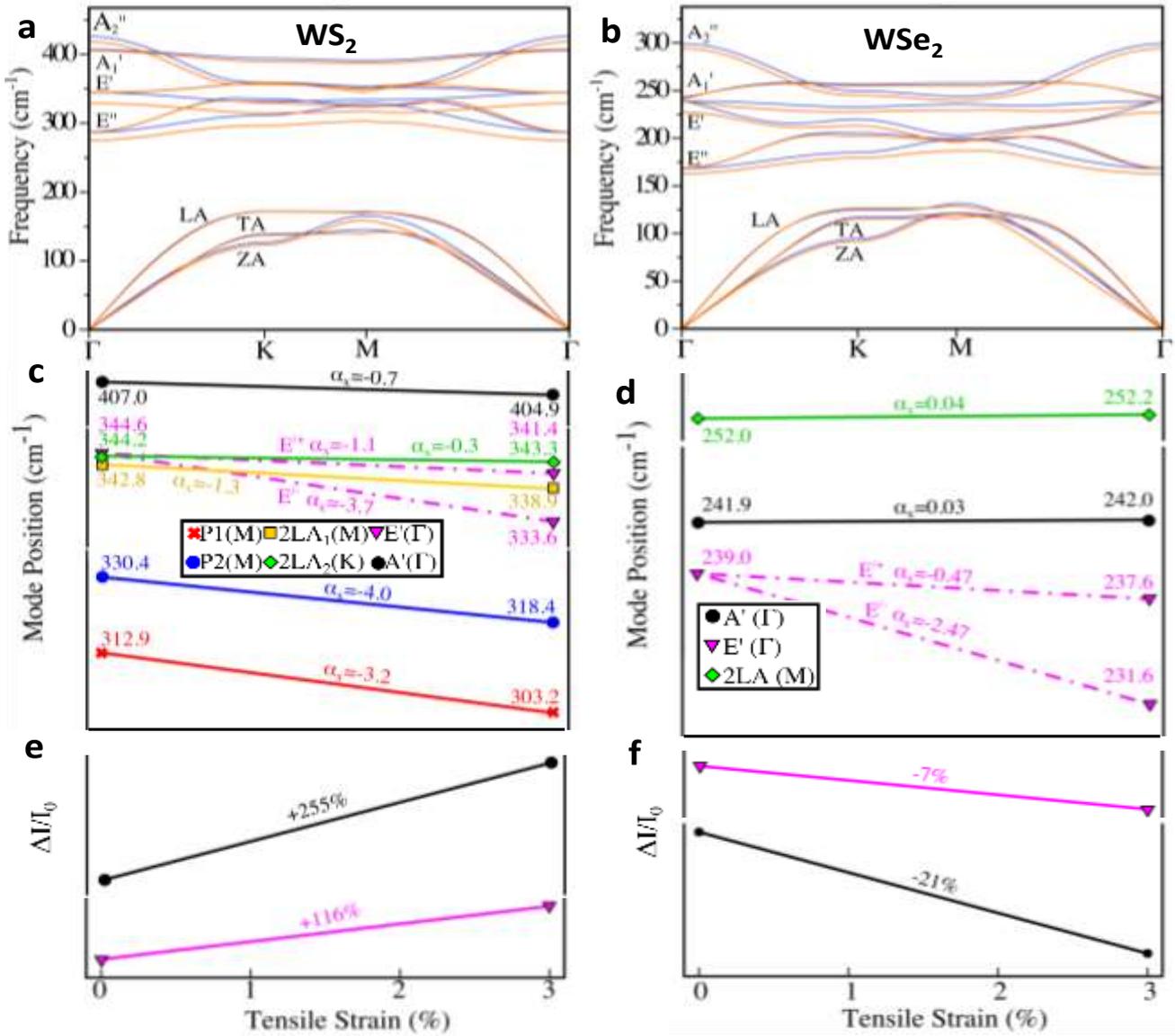

**Figure 4** – Theoretical results on the vibrational properties of WS2 and WSe2 under strain. (a,b)Phonon dispersion curves. The blue and orange curves indicate unstrained (0%) and strained (3%) systems, respectively with the strain applied along armchair direction. Similar results have been obtained for zigzag directions (not shown). (c, d)Variation in mode frequencies with applied strain. In (c) P1 and P2 are defined at the M-point; and 2LA1 and 2LA2 are the M-point and K-point respectively, apart from E' and A' at Γ-point. In (d) 2LA is defined as M-point, and A' and E' at Γ-point. α shows curve slopes. Polarizations for WS2 and WSe2 are set on A' at parallel; and E' at cross-polarized. (e, f)Relative intensity change of the Raman active modes at Γ-point

# Supporting Information

# Achieving Large, Tunable Strain in Monolayer Transition-Metal Dichalcogenides

Abdollah (Ali) M. Dadgar [1,2], Declan Scullion [3], Kyungnam Kang [4], Daniel Esposito [5], Eui-Hyeok Yang [4], Irving P. Herman [6], Marcos A. Pimenta [7], Elton-J. G. Santos [8], Abhay N. Pasupathy [2]

[1] Department of Mechanical Engineering, Columbia University, New York, NY 10027, USA

[2] Physics Department, Columbia University, New York, NY, 10027, USA

[3] School of Mathematics and Physics, Queens University, Belfast, BT7 1NN, UK

[4] Department of Mechanical Engineering, Stevens Institute of Technology, Castle Point on the Hudson, Hoboken, NJ, USA

[5] Department of Chemical Engineering, Columbia University - New York, NY 10027, USA

[6] Department of Applied Physics and Applied Mathematics, Columbia University, New York - NY 10027, USA

[7] Department of Physics, Universidade Federal de Minas Gerais (UFMG), Brazil

[8] School of Chemistry and Chemical Engineering, Queen's University - Belfast, BT9 5AL, UK

**Strain Method**

Our samples are monolayer flakes with a typical size of 10 μm. Direct stretching of the flake along the strain direction requires exceptionally precise control of the displacement, which is not easily achieved in a simple mechanical apparatus. We thus use an apparatus to bend the substrate in which the flakes are embedded as a simple way to achieve controllable strains. Areas above the neutral axis of the substrate undergo tensile strain while areas below the neutral axis experience compressive strain. It is typically assumed that the substrate bends in a circular arc upon bending. While reasonable at small strains, this approximation is incorrect at higher strains. In order to resolve this issue, we solve the Euler-Bernoulli's equation [45] for our experimental geometry to compute strain as a function of vertical displacement. We approximate the substrate to be a fully elastic, linear and isotropic material that obeys Hooke's Law.

Figure S1-a is the static free body diagram of the substrate under extra neutral axis bending. The ends of the beam (between A and B as well as between C and D in the figure S1-b) are cantilevers, with a center region between B and C undergoing pure bending. We first consider the center region and solve Euler-Bernoulli's equation for the vertical position of the substrate y as a function of the horizontal position x:

$$EI\frac{d^2y}{dx^2} = M = Pa \qquad (1)$$

where $E$ is Young's modulus of elasticity, $I$ is cross sectional moment of inertia, $M$ is bending moment and $P$ is the vertical load on the system. Since this region is under pure bending, the bending moment, $M = Pa$ is position independent (figure S1-c-right and S1-b-right). As a consequence, the strain in this region is also position independent. This property of the technique is highly desirable to achieve uniform strain on the sample and is the main reason for choosing this geometry. Solving this differential equation and applying boundary conditions, the final equation of the beam in the center region is:

$$EIy = \frac{1}{2}Pax^2 - \frac{1}{2}Pabx \qquad (2)$$

In the cantilever region, the bending moment is position dependent: $M = Px$ which is shown in figure S1-c,-left panel.

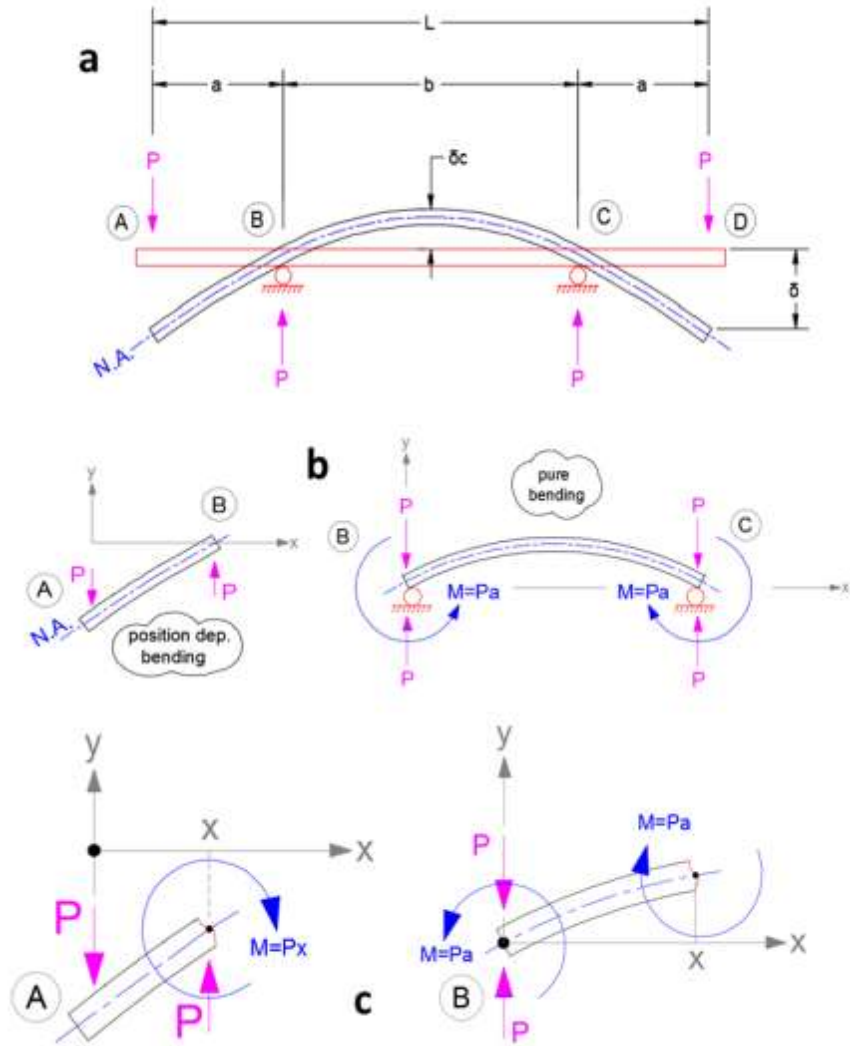

Figure S1 – **Statics equilibrium of beam under extra neutral axis bending**. (a) Statics of the whole beam under bending. (b) Beam is divided into two sections: (left) cantilever sections and (right) center section. (c) Details of loading at an arbitrary point of these regions.

Solving the Euler-Bernoulli equation and applying the appropriate boundary conditions gives the deflection equation in this region:

$$EIy = \frac{1}{6}P(x^3 - a^3) - \frac{1}{2}Pa(x-a)(a+b) \quad (3)$$

Using this function, we can now relate $\delta$ the vertical displacement at $x = 0$ to the other parameters of the beam:

$$\delta = \frac{1}{6}\frac{Pa^2}{EI}(2a + 3b) \quad (4)$$

Meanwhile, the maximum stress $\sigma$ in the center region can be calculated using Hooke's Law [1]: $\sigma = pat/2I$, where $t$ is the thickness of the substrate. Re-writing Hooke's Law for the uniaxial stress case, we will have the strain relation $\varepsilon = pat/2EI$. Combining this with equation (4), we get our desired relation – $\varepsilon$ as a function of beam geometry and vertical displacement $\delta$:

$$\varepsilon = \frac{3t\delta}{a(3b+2a)} \tag{5}$$

By choosing $a, b, t$ appropriately and controlling the vertical motion $\delta$ stepwise using a differential screw, we can apply linear stepwise strain to the flakes being studied.

One main remaining challenge though is to make a fully elastic springy material with larger elastic region which obeys that formula. We tried several annealing processes to re-crystallize polymer substrates and make them behave fully linear in elastic regime. We first measured glass transition temperature $T_g$ of different polymer materials and stacks. Then after encapsulating our semiconductor devices in our desired polymer, we heated it up to the annealing temperature $T_a = 0.97 T_g$. The optimized annealing time found to be $t_a = 60\ min$ for the substrates with $t = 0.5\ mm$ thicknesses. We found that applying pressure of $P_a = 3\ psi$ during annealing time, makes better re-crystallized substrates. Finally, very slow cooling with $r_a = 2\ °C/min$ rate, completes the substrate fabrication. These devices reveal excellent match with equation (5)

**Strength or Intensity of Raman Scattering**

The intensity of the Raman response for a given light polarization can be determined from the Raman tensor $\tilde{R}$ [46]:

$$I \propto |\hat{e}_s^T . \tilde{R} . \hat{e}_i|^2$$

where $\hat{e}_i$ and $\hat{e}_s$ are the polarization directions of the incident and scattered light's electric field respectively. A typical cross polarization geometry is illustrated in Figure S2.

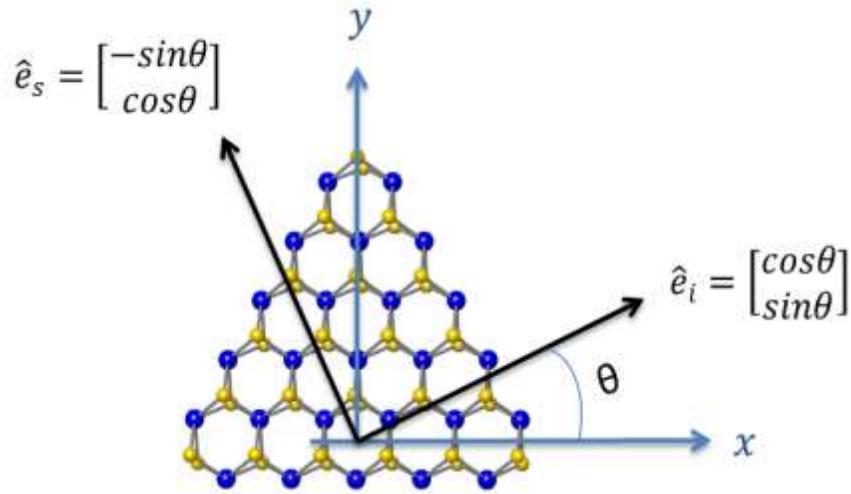

**Figure S2 – Directions of incident and scattered light with respect to crystal's Cartesian coordinate systems.** x- and y-directions represent zigzag and armchair directions respectively

When the incident laser light is applied in either the zigzag or armchair direction of a single layer orthorhombic crystal, the Raman tensor for the $A'$ mode is a diagonal matrix which can be written as follows:

$$R = \begin{bmatrix} a & 0 \\ 0 & b \end{bmatrix}$$

Consequently the Raman intensity of the $A'$ mode in the cross polarization geometry shown in figure S2 is:

$$I_{XY}^{\acute{A}} \propto \left| [\cos\theta \quad \sin\theta] \begin{bmatrix} a & 0 \\ 0 & b \end{bmatrix} \begin{bmatrix} -\sin\theta \\ \cos\theta \end{bmatrix} \right|^2 = |(b-a)\cos\theta\sin\theta|^2$$

This implies that when the incident light is along the crystal's principal directions, $\theta = 0°$ or $\theta = 90°$ (zigzag or armchair directions in figure S2), the $A'$ mode is not observed in the cross polarization geometry.

In the presence of larger strains the orthorhombic symmetry of the crystal is lowered to triclinic, in which the corresponding Raman tensor for the $A'$ mode is a non-diagonal matrix:

$$R = \begin{bmatrix} a & c \\ -c & b \end{bmatrix}$$

in this case the Raman intensity of the $A'$ mode is always nonzero :

$$I_{XY}^{\acute{A}} \propto \left| [cos\theta \quad sin\theta] \begin{bmatrix} a & c \\ -c & b \end{bmatrix} \begin{bmatrix} -sin\theta \\ cos\theta \end{bmatrix} \right|^2 = |(b-a)cos\theta sin\theta + c|^2 \neq 0$$

This might be the reason that we observe $A'$ mode at highly-strained WS$_2$ samples only. Subsequently, for WSe$_2$, there might be also possible contributions of $A'$ mode at higher strains but we could not observe because $A'$ and $E'$ modes are accidentally overlapped in this material.

**General Theoretical Methods:**

Theoretical calculations were performed using the DFT formalism as implemented in the Vienna ab initio simulation package (VASP)[47, 48]. For the calculations of phonon modes, both PHONON [49, 50] was used. A $2x2x1$ supercell was used for these calculations. This was increased to $3x3x1$ and there was no appreciable difference found in the Γ-point phonon frequencies. For the calculations of Raman band frequency, PHONON was used implementing the PEAD method. The PBE functional [51] was used along with the plane-wave cutoff of 800 eV combined with the projector-augmented wave (PAW) method [52, 53]. Atoms were allowed to relax under the conjugate-gradient algorithm until the forces acting on the atoms were less than $1x10-8\ eV/A$. The self-consistent field (SCF) convergence was also set to $1.0x10-8\ eV$. We employ an orthorhombic cell to apply uniaxial strain. Relaxed lattice constants were found to be $a = 3.187A, b = 5.52A$ for monolayer WS2 and $a = 3.319A, b = 5.750A$ for WSe2. A $20A$ vacuum space was used to restrict interactions between images. A $182x12x1$ gamma-centered k-grid was used to sample the Brillouin Zone for both systems.

**First-Principle Ab Initio Calculations**

Theoretical calculations were carried out on the n×n (n=1,2,3) orthorhombic cells for both WS$_2$ and WSe$_2$ (see Figure S3). The lattice constants were found to be a=3.187Å, b=5.520Å for WS2 and a=3.319Å,

b=5.750Å for WSe$_2$. A 25Å vacuum space was used in all calculations. Strain was applied along of the armchair direction. There was no observed geometric transition from an orthorhombic unit cell to a monoclinic unit cell after the strain was applied. Even though, the high magnitudes of strain (>3.0%) can incidentally change the local symmetry of the atoms from D3h to a lower-symmetry group.

At zero strain, under cross polarization, both the P1 and A' mode are suppressed. When strain is applied to the system (~3.0%), we see an increment of the intensity from zero to 5x10-6 a.u. for the $A'$ mode and 1x10-8 a.u. for the $P_1$ mode. The large difference in intensities between both modes (500 times) indicates that the $P_1$ mode is incredibly weak relative to $A'$, which is almost at the numerical accuracy of the calculation. As a matter of fact, just the $A'$ mode is experimentally observed at the high-strain regime (figure 3a in the main text), which fully corroborates the vibrational analysis performed. This suggests that large strain can indeed change the selection rules for the modes, even though their weak intensities may not be measured within the limit of resolution of the analyzer.

**Growth of Monolayer WS2 on a SiO2 Substrates**

The sample was comprised of a tungsten source carrier chip (5nm WO3 thin film on 90 nm SiO2) and bare SiO2/Si substrate (90 nm thick SiO2, WRS materials). Tungsten oxide (WO3, 99.99%, Kurt J. Lesker) was deposited on SiO2 via electron beam evaporation. The tungsten source chip was covered, in face-to-face contact, by a bare SiO2/Si substrate as the growth substrate. The sample was loaded into the center of a 3" diameter and 1 m long quartz tube (MTI Corp.), and a ceramic boat with 0.8 g of sulfur powder (99.98%, Sigma-Aldrich) was located upstream in the quartz tube. After loading, the ambient gas of the tube was purged out via mechanical pump to the base pressure of 400 mTorr. The furnace was heated to 700 ˚C at a 20 ˚C/min ramping rate and then to 900 ˚C at 5 ˚C/min. 60 sccm of Ar gas (5.0 UH purity, Praxair) was introduced at 150 ˚C (increasing temperature) to reduce moisture inside of the tube and closed at 600 ˚C (decreasing temperature). Hydrogen (40 sccm, 5.0 UH purity, Praxair) gas was supplied to improve WO3 reduction from 700 ˚C (increasing temperature) to 600 ˚C (decreasing temperature). The growth pressure was 7 Torr. After 25 min at 900 ˚C, the furnace was cooled down to room temperature.

**Growth of Monolayer WSe2 on a SiO2 Substrates**

The sample was comprised of a tungsten source carrier chip (5nm WO3 thin film on 90 nm SiO2) and bare SiO2/Si substrate (90 nm thick SiO2, WRS materials). Tungsten oxide (WO3, 99.99%, Kurt J. Lesker) was deposited on SiO2 via electron beam evaporation. The tungsten source chip was covered, in face-to-face contact, by a bare SiO2/Si substrate as the growth substrate. The sample was loaded into the center of a 2" diameter and 24" long quartz tube (MTI Corp.), and a ceramic boat with 0.8 g of selenium powder (99.99%, Sigma-Aldrich) was located upstream in the quartz tube. After loading, the ambient gas of the tube was purged out via mechanical pump to the base pressure of 100 mTorr. The furnace was heated to 700 ˚C at a 20 ˚C/min ramping rate and then to 900 ˚C at 5 ˚C/min. 20 sccm of Ar gas (5.0 UH purity, Praxair) was introduced at 150 ˚C (increasing temperature) to reduce moisture inside of the tube and closed at 600 ˚C (decreasing temperature). Hydrogen (10 sccm, 5.0 UH purity, Praxair) gas was supplied to improve WO3 reduction from 700 ˚C (increasing temperature) to 600 ˚C (decreasing temperature). The growth pressure was 1 Torr. After 25 min at 900 ˚C, the furnace was cooled down to room temperature.

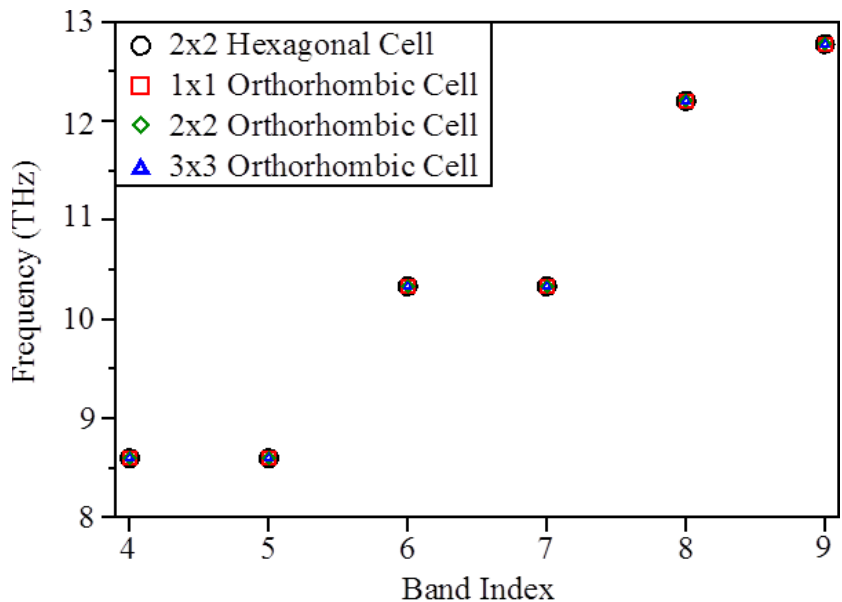

**Figure S3 – Convergence of phonon frequencies with respect to the size of the supercell.** As it can be seen, there is no change in the phonon frequency while changing the supercell size.

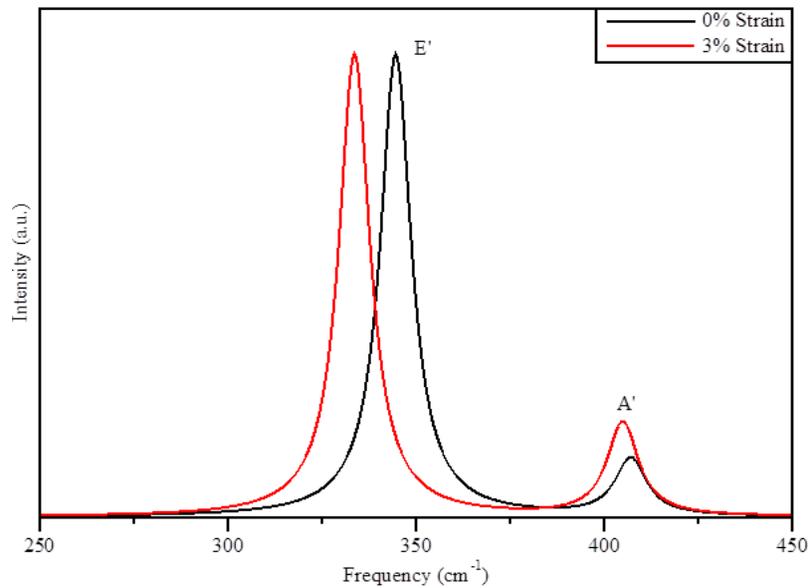

**Figure S4 – Raman peaks for WS2 generated using a Gaussian smearing with the A' and E' modes labelled.** As it can be seen both modes decrease in frequency with applied strain whilst the intensity increases.

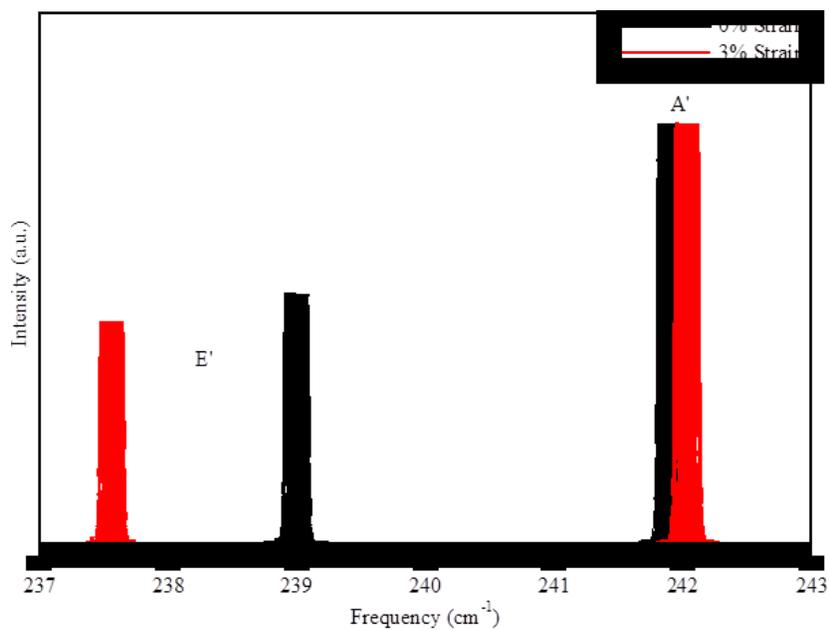

**Figure S5 – Raman peaks for WSe2 generated using a Gaussian smearing with the $A'$ and $E'$ modes labelled.** It can be seen that the e' mode decreases in frequency with applied strain whereas the A' mode increases in frequency. Both modes decrease in intensity with applied strain.

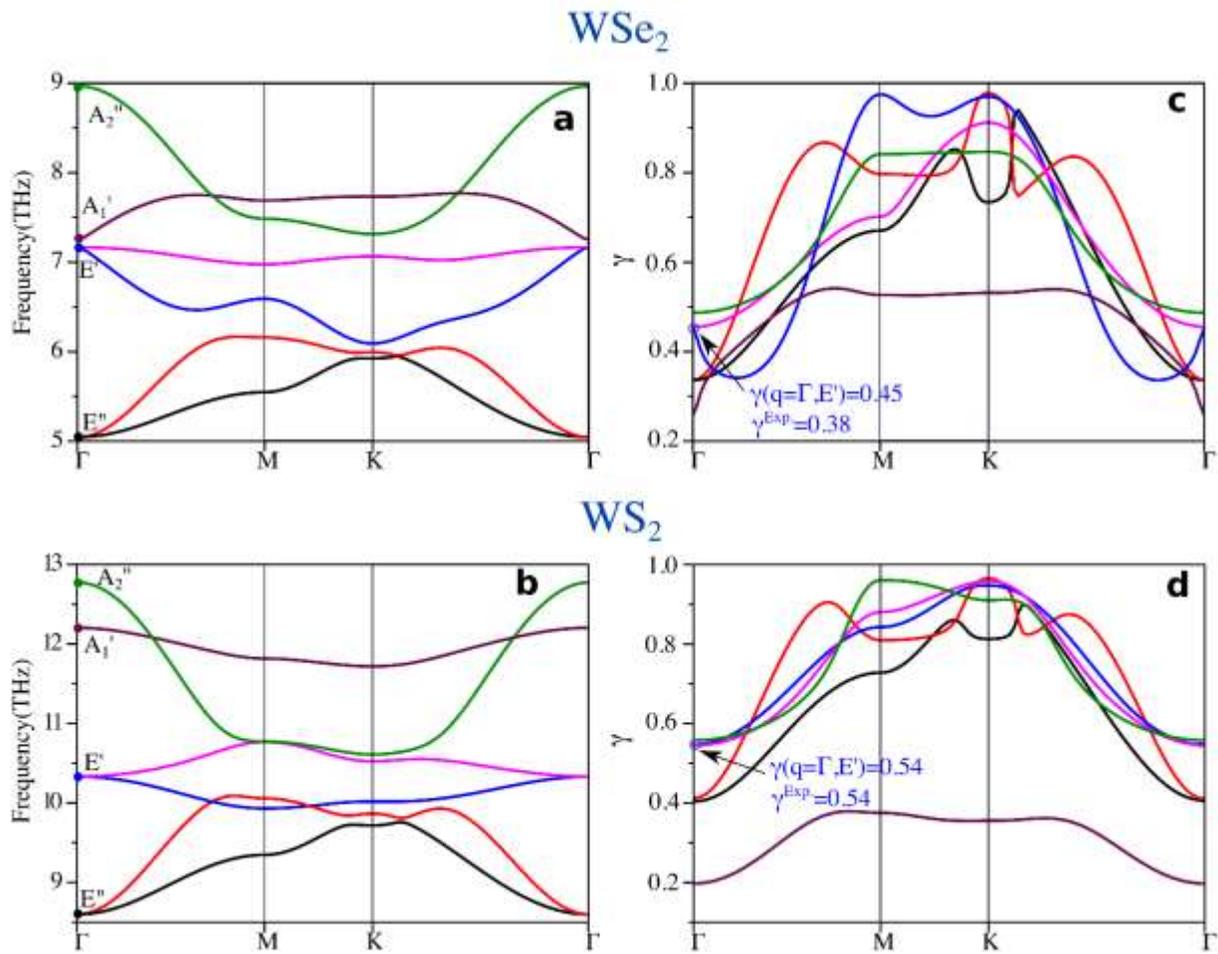

**Figure S6 – (a,b) Phonon dispersion branches for optical modes ($A_2''$, $A_1'$, $E'$, $E''$) and their respective (c,d) Grüneisen parameters γ calculated along of Γ−M−K−Γ path for monolayer WSe2 and WS2.** Modes are labeled according to their symmetries at Γ−point (filled dots) in a and b. The different colors correspond to different phonon branches, which follow the same labeling for γ. At Γ−point, the calculated values of the Grüneisen parameters are γ=0.45 and γ=0.54 for WSe2 and WS2, respectively. This is in remarkable agreement with the experimental results γExp.=0.38 (WSe$_2$) and γ Exp.=0.54 (WS$_2$).